\documentclass[a4paper, 11pt]{article}
\usepackage{graphicx}
\usepackage{amsmath,amssymb}
\usepackage{amsfonts}
\usepackage{color}
\usepackage[center, small, sl]{caption}

\makeatletter

\title{Racetrack FFAG muon decay ring for\\
 nuSTORM with triplet focusing}
\author{J.-B. Lagrange$^{(1,2)}$, R.B. Appleby$^{(3)}$, J.M. Garland$^{(3)}$, J. Pasternak$^{(2,4)}$, S. Tygier$^{(3)}$ \\
(1) FNAL, IL, USA\\(2) Imperial College, UK\\(3) University of Manchester and the Cockcroft Institute, UK\\(4) ISIS-RAL-STFC, UK}
\date{}

\begin{document}

\maketitle

\begin{abstract}
The neutrino beam produced from muons decaying in a storage ring would be an ideal tool for precise neutrino cross section measurements and the search for sterile neutrinos due to its precisely known flavour content and spectrum. In the proposed nuSTORM facility, pions would be directly injected into a racetrack storage ring, where the circulating muon beam would be captured. In this paper we show that a muon decay ring based on a racetrack scaling FFAG (Fixed Field Alternating Gradient) with triplet focusing structures is a very promising option with potential advantages over the FODO based solution. 
We discuss the ring concept, machine parameters, linear optics design, beam dynamics and the injection system.
\end{abstract}

\section{Introduction}
Using a decaying muon beam to produce a neutrino beam with a defined spectrum and flux composition is a well-established idea. This concept was developed into the Neutrino Factory facility proposal, which was then addressed in several dedicated research and development studies culminating in the International Design Study for the Neutrino Factory (IDS-NF)~\cite{idsnf}. The Neutrino Factory consists of a high power proton driver~\cite{proton_driver}, the output of which is directed towards a pion production target~\cite{target}; a decay channel, where the muon beam is formed; the muon front end~\cite{front_end}, where the beam is prepared for the acceleration~\cite{nf_acc} and the muon accelerator to boost the energy to the required value. The muon beam is then injected into the decay ring~\cite{nf_decay_ring}, with straight sections pointing towards near and far detectors producing neutrino beams for both interaction and oscillation physics. Although it has been shown~\cite{idsnf} that such a facility will have superior physics potentials for the leptonic CP violation searches with respect to a conventional neutrino beam facility based on pion decay, it requires the construction of many new accelerator components that do not exist at present. 

In order to allow for the start of neutrino physics experiments based on muon decay using conventional accelerator technology, the neutrinos from STORed Muon beam (nuSTORM) project was proposed~\cite{nustorm-adey}. The main goal of nuSTORM is to precisely study neutrino interactions for electron and muon neutrinos and their antiparticles, but the facility could also contribute to sterile neutrino searches, and serve as a proof of principle for the Neutrino Factory concept. In nuSTORM high energy pions produced at the target are first focused with a magnetic horn~\cite{nustorm-horn-ao}, and directly injected into the ring after passing through a short transfer line equipped with a chicane to select the charge of the beam. Once in the ring, decaying pions will form the muon beam. A fraction of the muon beam with momentum lower than the injected parent pions will be stored in the ring and a fraction with similar or larger momentum will be extracted with a mirror system of the injection at the end of the long straight section to reduce beam loss, and to avoid activation in the arc. The extracted beam may also be used for accelerator research and development studies for future muon accelerators, which may serve as another application for nuSTORM. The design and performance of the decay ring is tightly linked to the neutrino physics reach of the experiment, with intrinsic design challenges arising from the large range of beam momenta in the ring. 

At the present time there are two options under study for the design of the decay ring. The first one is the FODO solution with large bore conventional quadrupoles with alternating gradients \cite{ao-thesis} in the long straight sections and with lattice based on separate function magnets in the arcs. This solution provides excellent performance with respect to the transverse acceptance, but it has very limited longitudinal acceptance~\cite{ao-thesis} due to the natural chromaticity, resulting directly from alternating gradient conventional magnet approach. The second solution is the use of recent developments in Fixed Field Alternating Gradient (FFAG) accelerators. In these machines, which come in scaling and non-scaling flavours, large aperture non-linear magnets allow the beam to move through the aperture at varying momenta, with a constant betatron tune in the case of a scaling FFAG. The advantage of such a lattice is the large momentum acceptance together with a large transverse acceptance, thus increasing the number of stored muons in the ring. In the present case, we propose to double the momentum acceptance compare to the FODO solution. The design is realised by keeping the ring zero-chromatic over the whole momentum range, and choosing the tune point far from harmful resonances. Moreover, a racetrack shape is now possible~\cite{norma}, while keeping the ring zero-chromatic for a large momentum range, thanks to use of straight scaling FFAG cells.

The constant betatron tune with momentum provides a strong constraint on the fields in a scaling FFAG. In the arcs, the vertical magnetic field $B_{\mathrm{az}}$ in the median plane produced by the combined-function magnets follows the circular scaling law
\begin{displaymath}
B_{\mathrm{az}}=B_{\mathrm{0az}} \left( \frac{r}{r_0}\right)^k \mathcal{F},
\end{displaymath}
with $r$ the radius in the polar coordinate, $k$ the constant geometrical field index, $ \mathcal{F}$ an arbitrary azimuthal function used to represent the position of the magnets and the fringe fields, and $B_{\mathrm{0az}}=B_{\mathrm{az}}(r_0)$~\cite{symon-1956}. This type of magnet has been successfully built for several machines in the past few decades~\cite{mura1, mura2, pop, kek, erit, prism}. In the straight sections, the vertical magnetic field $B_{\mathrm{sz}}$ in the median plane produced by the combined-function magnets follows the straight scaling law
\begin{displaymath}
B_{\mathrm{sz}}=B_{\mathrm{0sz}} e^{m(x-x_0)}  \mathcal{F},
\end{displaymath}
with $x$ the horizontal cartesian coordinate, $m$ the constant normalized field gradient, and $B_{\mathrm{0sz}}=B_{\mathrm{sz}}(x_0)$~\cite{lagrange-nim-2012}. 
The beam orbit oscillations through scaling magnets in this type of straight produce a characteristic periodic beam angle oscillation known as a scallop angle, which ultimately limits the achievable neutrino flux. Straight scaling FFAG magnets have been successfully demonstrated experimentally~\cite{lagrange-nim-2012}.

In this paper we show that a scaling FFAG in a racetrack configuration, with straight sections and arcs built from magnets fulfilling these scaling laws, can provide a muon storage ring which 
meets the performance requirements of nuSTORM. The ring consists of straight sections matched with the regular scaling FFAG arcs using circular FFAG matching cells, which allow incorporating 
dispersion adjustment. We show the scallop angle can be optimised and a good long term dynamic behaviour can be established to maximise the neutrino flux. As the muon beam intensity is proportional to the momentum acceptance, the broad acceptance of FFAG rings offers a significant advantage. We show here, for the first time, that the combination of large momentum bandwidth and optimised optics can provide a path to achieve a sufficient neutrino flux for nuSTORM and related muon decay experiments. 

In section~\ref{sec:design}, we show the design of the FFAG ring solution, the lattice parameters and the magnetic field distributions. We also present details on the optics matching procedure. 
Then in section~\ref{sec:perf} we present the performance of the lattice, highlighting the long-term stability and the delivered flux. 
We show that the flux requirements of the nuSTORM storage ring can be met with a FFAG solution, with acceptable dynamic aperture and beam dynamics, bringing a complete and feasible design of nuSTORM a significant step closer. Finally in section~\ref{sec:summ} we draw our conclusions. 

\section{The FFAG ring parameters and design}
\label{sec:design}

In this section we motivate and describe the parameters and design of the FFAG ring, including definition of the fields and optical properties. We also discuss the constraints on the design from the perspective of flux optimisation. 

The main constraint of the FFAG solution is to keep the scallop angle of the reference trajectories in the straight section as small as possible. A large scallop angle, the deviation of the beam from a straight line, will alter the spectrum and overall value of the neutrino flux. This goal was addressed with either alternating gradient doublet or triplet lattice cells providing some optimisation. The advantage of a triplet over a doublet layout is to constrain the scallop part to a small portion of the cell. Dispersion has to be kept small in the muon production straight to keep a good muon capture rate. Since the central momentum of the injected pion is different from the central momentum of the circulating beam, the horizontal position of the reference trajectories for the two of them must not be too far away from each other so that the muon obtained from pion decay is within the acceptance of the circulating beam. However, the dispersion has to be large where the beam is injected to provide the necessary beam separation, so a dispersion matched section is necessary to accommodate the two constraints. A dispersion suppressor is then introduced in the arcs. The concept of a dispersion suppressor in scaling FFAG is to induce a betatron oscillation of the reference trajectories around the periodic trajectory of the matching cell for momenta that is different to the arbitrary-chosen matched one. The phase advance of the dispersion suppressor section is $180^{\circ}$, allowing half of a complete oscillation and a matching of the reference trajectories.

In order to match the horizontal beta-functions, the horizontal phase advance of the arc part is an integer number of $\pi$, and the vertical beta-functions are matched by adjusting the ratio of magnetic fields in the magnets, the so called F/D ratios. Superconducting magnets are considered in arc section to keep it compact and to have a large ratio of the production straight length to the circumference. While in the straight section, to drive down the manufacturing cost, room temperature magnets are preferred. However, the use of super-ferric magnets in the straight sections are also being considered due to lower power consumption. Lattice parameters are summarized in Table~\ref{ffag-general}.
\begin{table}[!h]
    \centering
\begin{tabular}{lcc}
\hline
\hline
Total circumference & 510~m\\
Length of one straight section & 180~m\\
One straight section/circumference ratio & 35\%\\
Central momentum & 3.8~GeV/c\\
Momentum acceptance & $\pm 16\%$\\
Max. scallop angle & 24~mrad\\
\hline
Number of cells in the ring: &\\
Straight cells & 36 \\
Arc matching cells & 8\\
Arc cells & 8\\
Matching momentum $p_0$ & 3.648~GeV/c\\
Minimum momentum $p_{min}$ & 3.192~GeV/c\\
Maximum momentum $p_{max}$ & 4.408~GeV/c\\
Ring tune point (H/V) at $p_0$ &  (6.91, 3.69)\\
\hline
\hline
\end{tabular} 
 \caption{Parameters of the ring lattice.}
 \label{ffag-general}
\end{table}

 \begin{figure}[h!]
	\begin{center}
		\includegraphics[width=15cm]{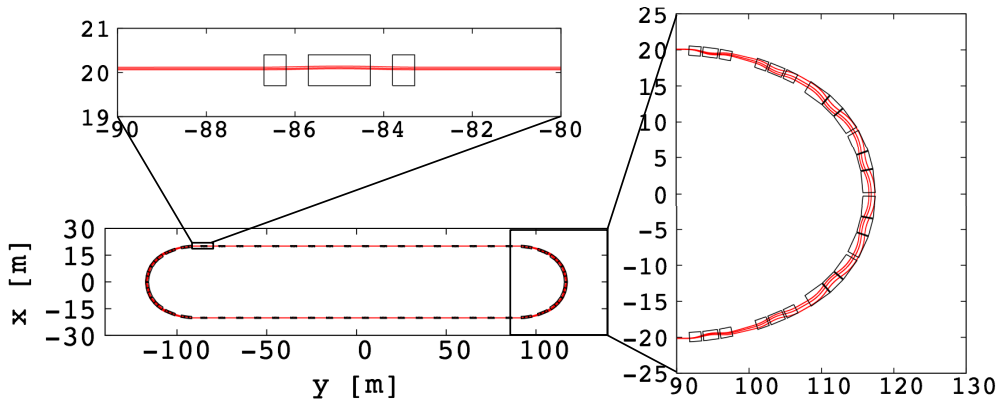}
		\caption{Top view of the racetrack FFAG lattice (bottom left scheme). The top left scheme shows a zoom on the straight section and the right scheme a zoom on the arc section. $p_0$, $p_{min}$, and $p_{max}$ muon closed orbits are shown in red. Effective field boundaries with collimators are shown in black.}
		\label{single-traj}
	\end{center}
\end{figure}

\begin{figure}[h!]
   \centering
    \includegraphics[width=7cm]{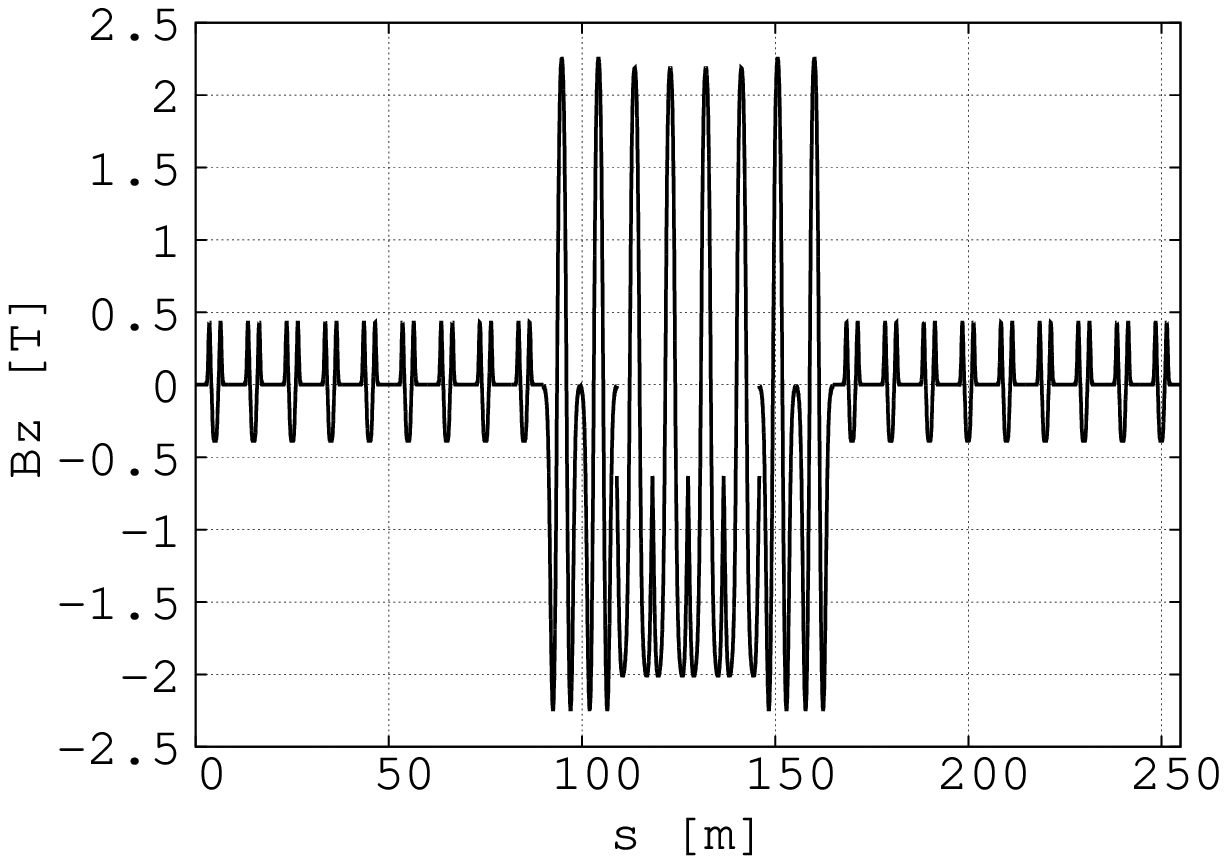}
       \includegraphics[width=7cm]{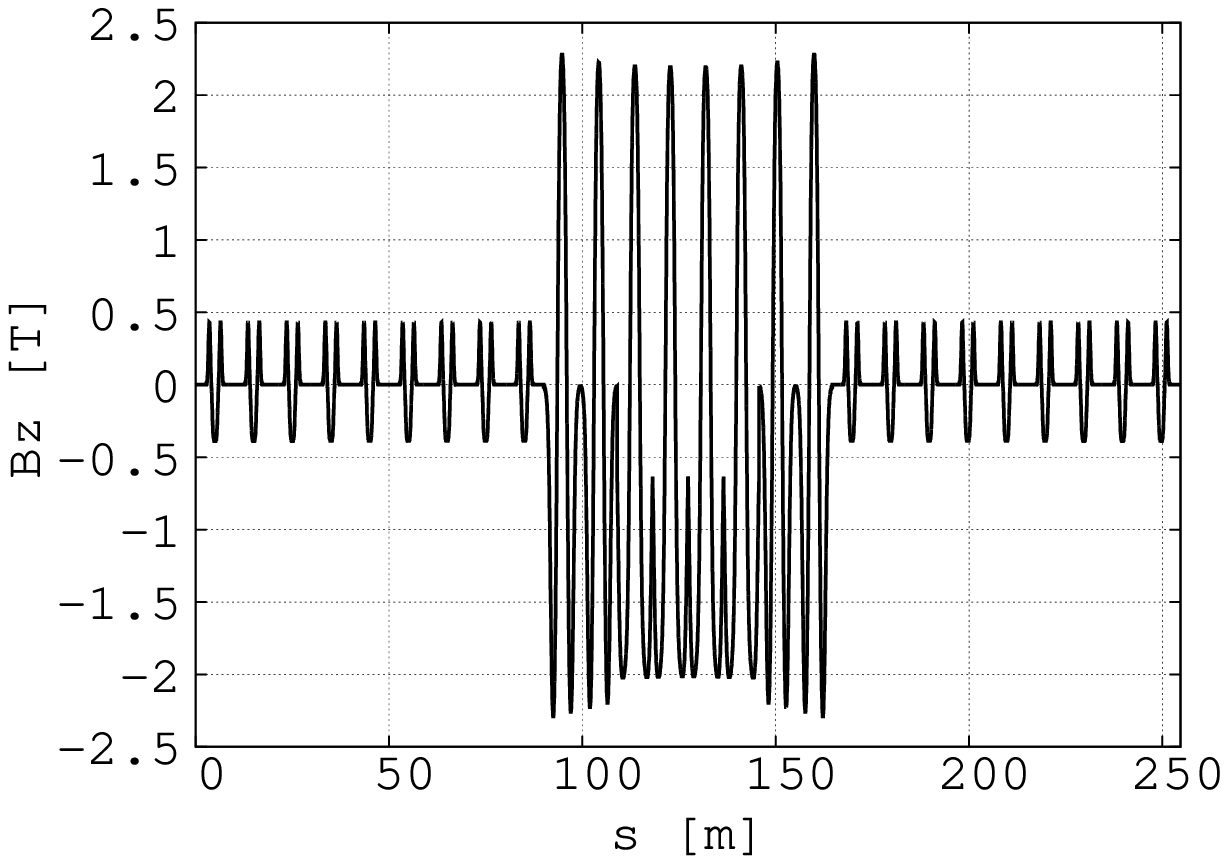}
       \\
        \includegraphics[width=7cm]{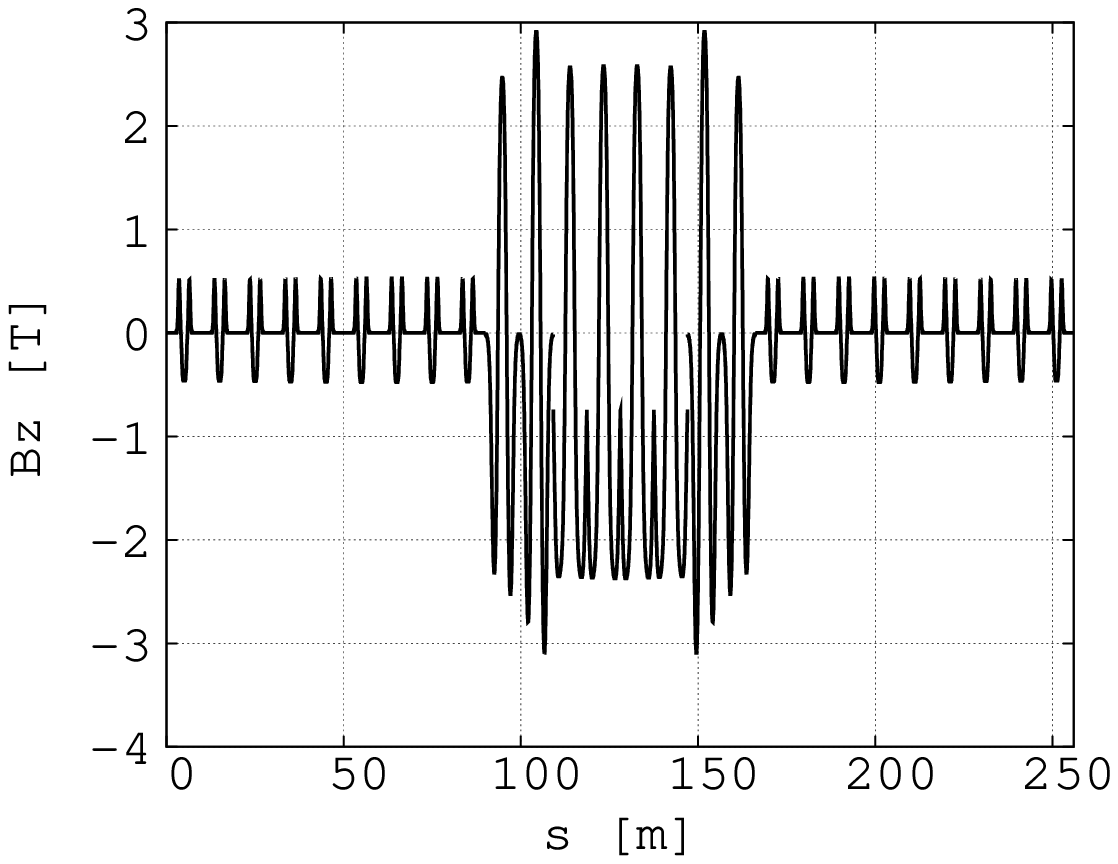}
\caption{The vertical magnetic field for $p_{0}$ muon closed orbits in half of the racetrack FFAG ring, centred on the arc part in JBT (top left) and in PyZgoubi (top right), and for $p_{max}$ in JBT (bottom).}
\label{bz}
\end{figure}

By using FFAG straight scaling triplet cells we have a achieved a maximum scallop angle of 24~mrad with a packing factor of 24\%, while the rest of the straight section does not have any scallop.
This allows optimising the deliverable flux. The machine layout resulting from this analysis is shown in Fig.~\ref{single-traj}, which shows the racetrack configuration in the lower-left, a zoom of the straight triplets in the upper left and a zoom on the arc in the right part of the figure. The corresponding vertical magnetic field for closed orbits with mean, $p_0$ and maximum, $p_{max}$, muon momentum are presented in Fig.~\ref{bz}, top left-hand plot and bottom-hand plot respectively. The maximum magnetic field stays in the normal conducting range in the straight section, while it reaches $\approx 3$~T in the arcs.

\begin{figure}[h!]
   \centering
    \includegraphics[width=15cm]{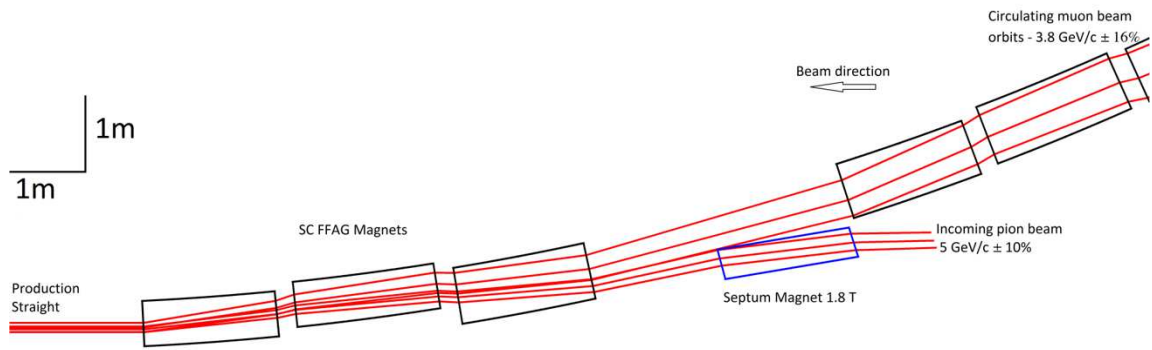}
\caption{Scheme of the injection for the FFAG solution.}
\label{injection}
\end{figure}

The so-called stochastic injection~\cite{neufer} is the preferred scheme in both the FODO and the FFAG solutions. Its principle is based on the use of the difference in momentum for the injected pions and circulating muon beam to inject in the ring without kicker. The incoming pion beam momentum is centred on 5~GeV/c, and the resulting muon decay is centred on 3.8~GeV/c. This difference in rigidity (equivalent to momentum) gives the possibility to have different orbits and sufficient separation between the injected beam and the circulating beam, providing the injected area has dispersion.
It has several advantages: the circumference of the ring is not constrained by the rise-time of the kicker, allowing it to be compact, it is cheaper, since it removes the need to have a pion decay channel, and finally it gives access to neutrino flux from pion decay in the production straight. This injection scheme can be adopted in the FFAG case by using the large dispersion in 
the matching section to inject the pions in the ring with a septum placed in the 2.6~m drift space designed for this purpose. A minimum dispersion of 1.34~m is required at the septum junction to 
have sufficient separation (~3~cm) between the incoming pion and circulating muon beams both with large momentum spread. The injection scheme is presented in Fig.~\ref{injection}.

\section{Performance}
\label{sec:perf}

In this section we study the beam dynamical properties of the lattice, including the optical functions, the long-term particle stability and the expected flux. The particle motion is performed with two techniques. The first is a stepwise tracking using Runge-Kutta integration, tracking through the fields shown in figure~\ref{bz}, with Enge-type fringe fields. This tracking model is denoted JBT. To check the validity of the long-term tracking simulations, the lattice has been implemented in the code PyZgoubi \cite{pyzgoubi} which uses the Zgoubi tracking engine \cite{zgoubi}. In the PyZgoubi simulation, the step size and fringe field parameters are the same as in the Runge-Kutta simulation. The straight scaling FFAG field law is implemented in PyZgoubi using a multipole fit of the magnetic field up to the decapole.

The parameters of the tracking model are summarized in Table~\ref{tracking-para}.
 \begin{table}[!h]
    \centering
\begin{tabular}{lcc}
\hline
\hline
Field type & Field model\\
Fringe field type & Enge\\
Interpolation off the mid-plane & 4$^{th}$ order\\
Step size & 5~mm\\
Particle & muon $\mu^+$\\
\hline
\hline
\end{tabular} 
 \caption{Parameters used in the tracking studies.}
 \label{tracking-para}
\end{table}

 \begin{figure}[h!]
	\begin{center}
		\includegraphics[width=12cm]{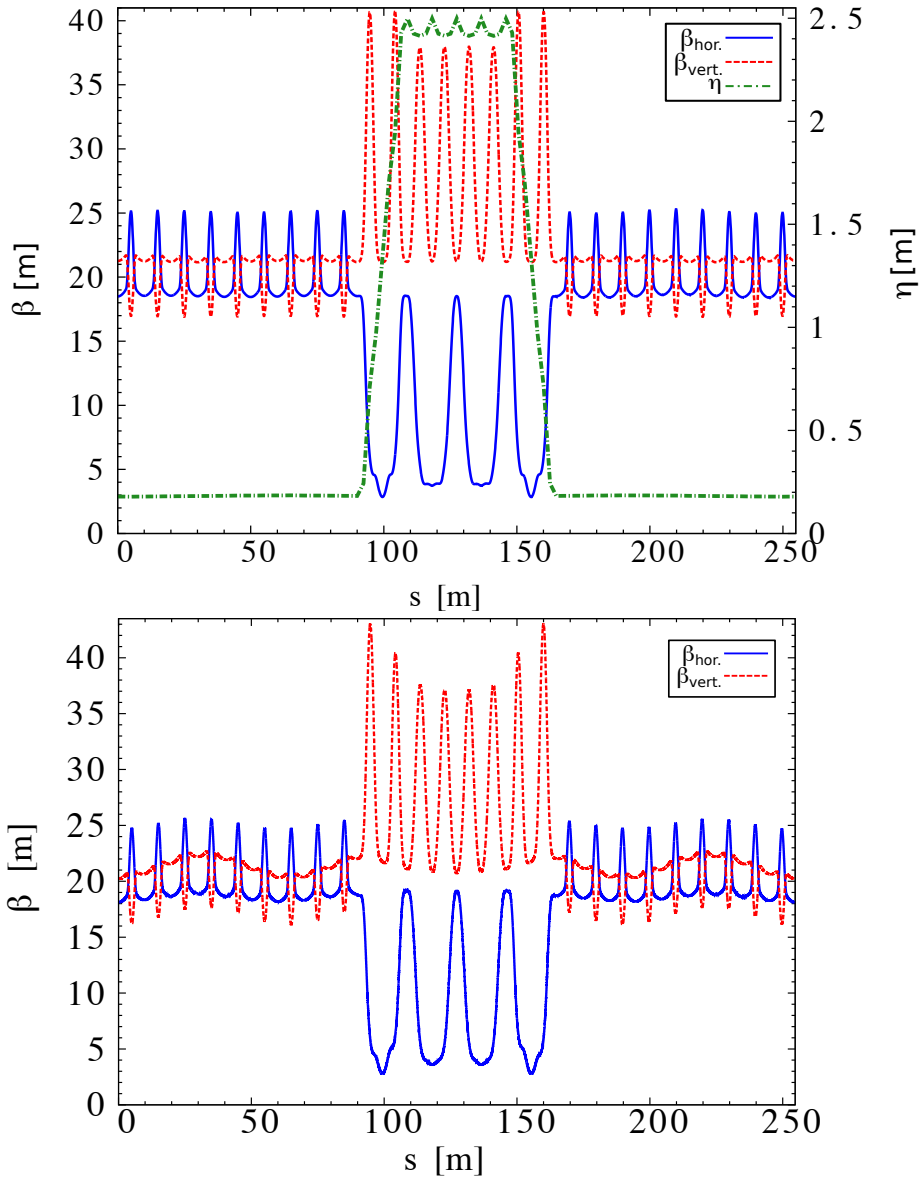}
		\caption{Horizontal (plain blue), vertical (dotted red) periodic beta-functions of half of the ring for $p_{0}$ in JBT (top) and in pyZgoubi (bottom). The dispersion (mixed green line) is also added in the top plot. The plots are centred on the arc part, that starts at $s=90$\,m and finishes at $s=165$\,m.}
		\label{beta}
	\end{center}
\end{figure}

\begin{figure}[h!]
	\begin{center}
		\includegraphics[width=12cm]{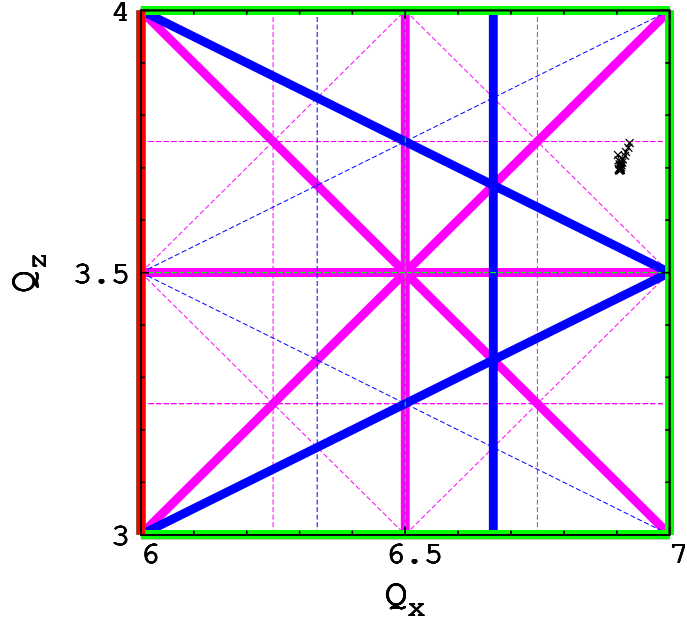}
		\caption{Tune diagram for momenta $\pm16\%$ around 3.8~GeV/c. Integer (red), half-integer (green), third integer (blue) and fourth integer (purple) normal resonances are plotted. Structural resonances are in bold. The very small tune shift demonstrates almost perfect zero-chromaticity.}
		\label{tunediag}
	\end{center}
\end{figure}

The variation of the excursion of the beam in the ring can be seen in the right-hand side of Fig.~\ref{single-traj} where closed orbits of the reference momentum $p_{\mathrm{0}}$, the minimum momentum $p_{\mathrm{min}}$, and the maximum momentum $p_{\mathrm{max}}$ particles are shown. The lattice functions dispersion and beta-functions at $p_{0}$ are shown in Fig.~\ref{beta} (top plot), showing the periodic oscillations in the straight sections and the arc. The maximum beta-functions in the ring are around 40m. The zero-chromatic behaviour of the ring lattice can be seen in Fig.~\ref{tunediag}, which shows the horizontal and vertical betatron tune space, with the tune for $\pm$ 16\% around the central momentum of 3.8 GeV/c together with the dominant resonance lines overlaid. The plot shows the possible tune shift for the full momentum band of the beam. 

\begin{figure}[h!]
   \begin{minipage}[b]{.53\linewidth}
   \centering
       \includegraphics[width=8.5cm]{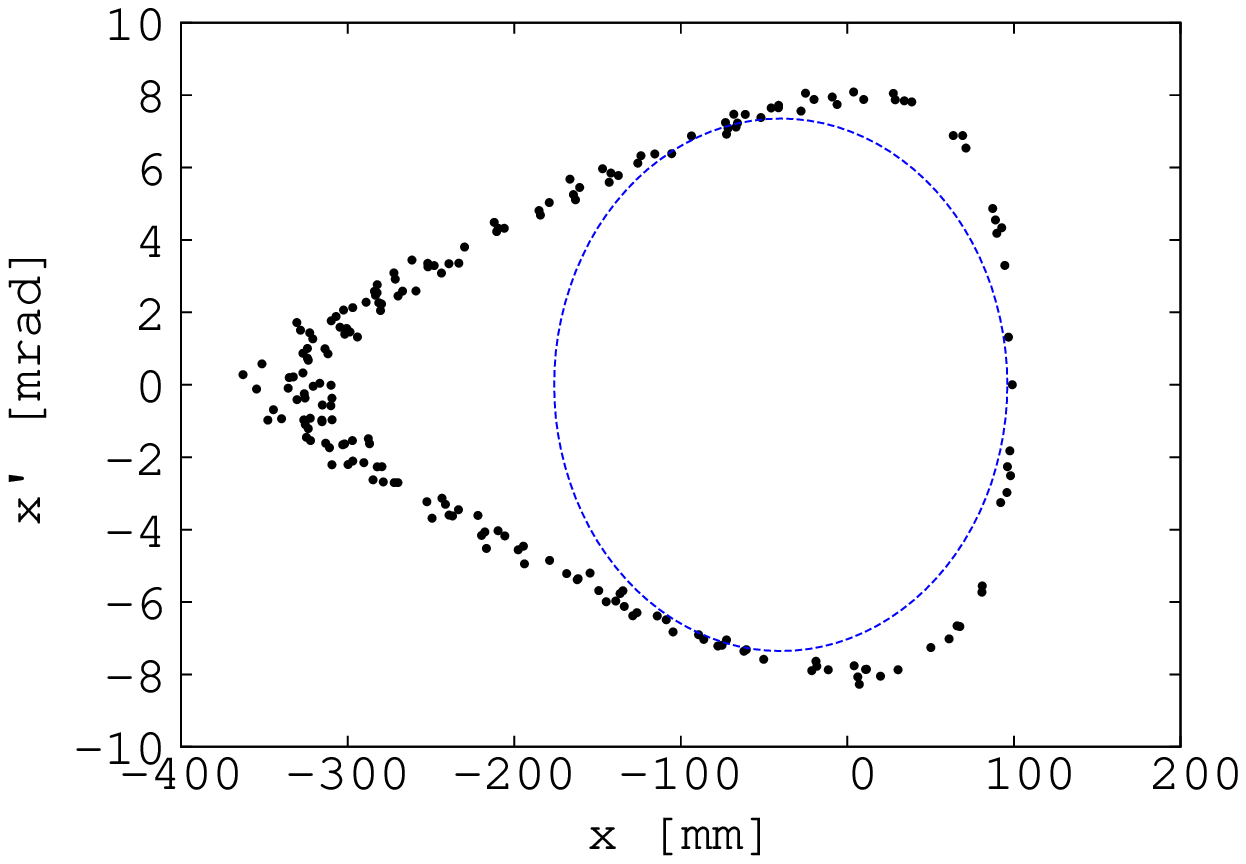}
	\caption{Stable motion in the horizontal Poincare map for maximum initial amplitude over 100 turns for $p_0$ in JBT. The ellipse shows a 1~mm.rad unnormalized emittance.}
	\label{poincarrex-single}
	 \end{minipage} \hfill
   \begin{minipage}[b]{.53\linewidth}
   \centering
    	\includegraphics[width=8.5cm]{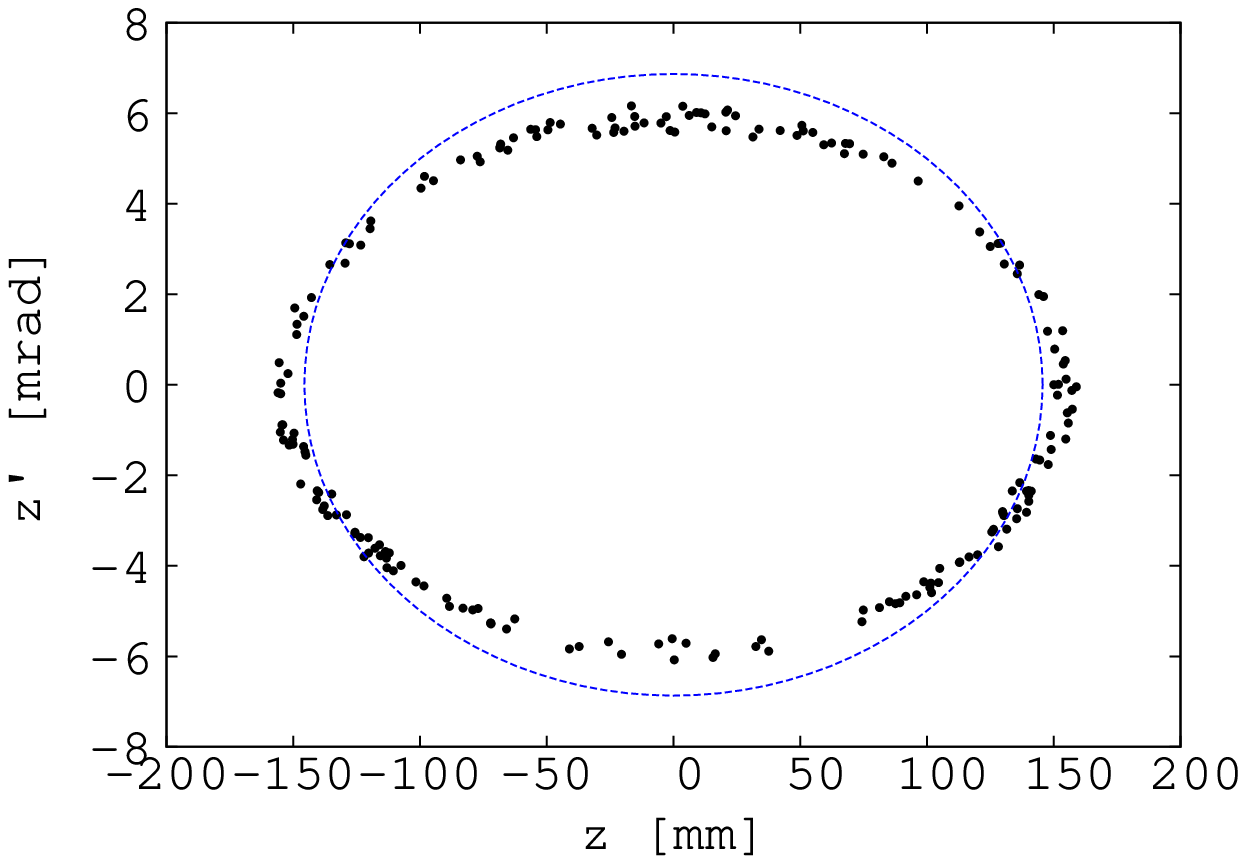}
	\caption{Stable motion in the vertical Poincare map for maximum initial amplitude over 100 turns for $p_0$ in JBT. The ellipse shows a 1~mm.rad unnormalized emittance.}
	\label{poincarrez-single}
   \end{minipage}
\end{figure}

We now compute the long term motion and the associated transverse acceptance using JBT. 
The transverse acceptance in both planes is studied by tracking over 100 turns a particle with a displacement off the closed orbit and a small deviation in the other transverse direction (1~mm). The regions drawn by the particle with the largest initial stable amplitude in the middle of the straight section in the horizontal and vertical phase spaces are presented in Fig.~\ref{poincarrex-single} and~\ref{poincarrez-single}, respectively. Acceptance of around 1~mm.rad is achieved in both planes.

\begin{figure}[h!]
   \begin{minipage}[b]{.53\linewidth}
   \centering
       \includegraphics[width=8.5cm]{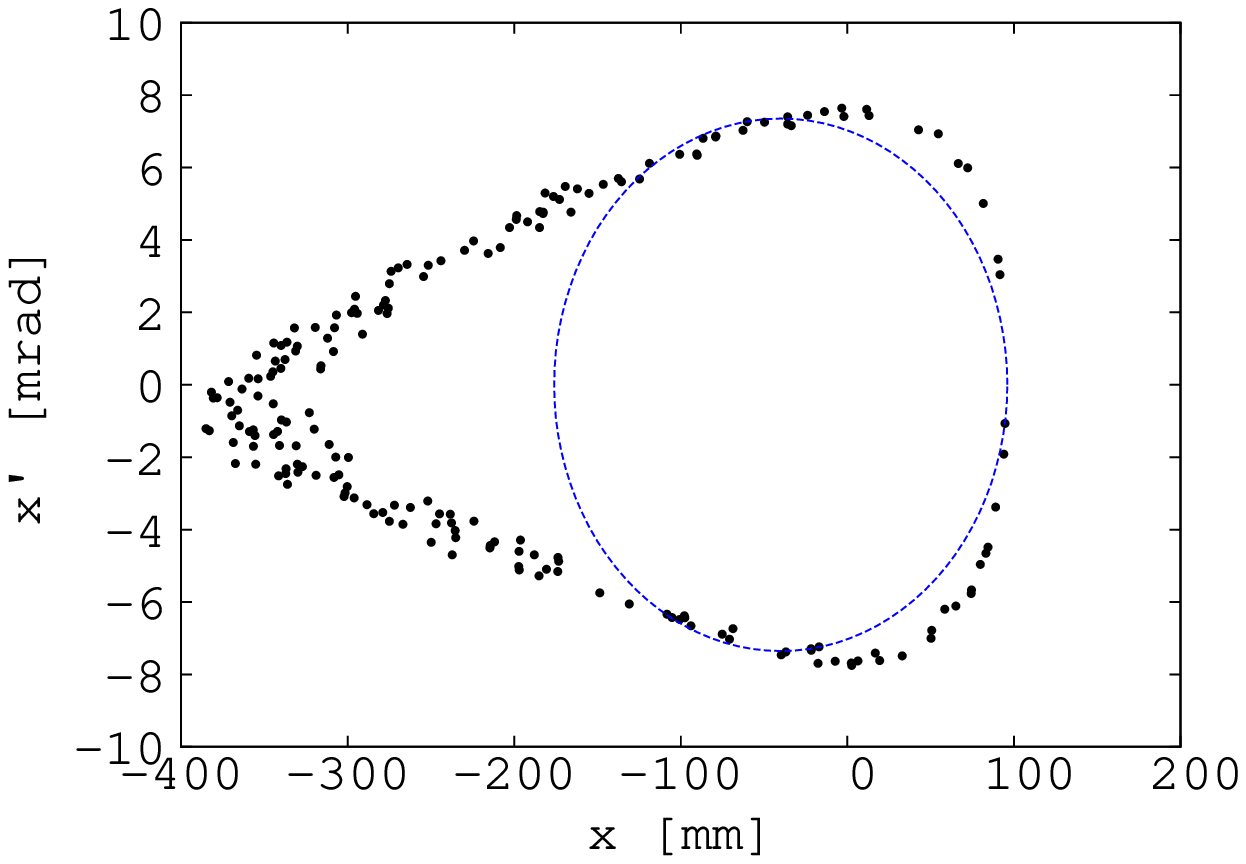}
	\caption{Stable motion in the horizontal Poincare map for maximum initial amplitude over 100 turns for $p_0$, in PyZgoubi. The ellipse shows a 1~mm.rad unnormalized emittance.}
	\label{poincarrex-pyzgoubi}
	 \end{minipage} \hfill
   \begin{minipage}[b]{.53\linewidth}
   \centering
    	\includegraphics[width=8.5cm]{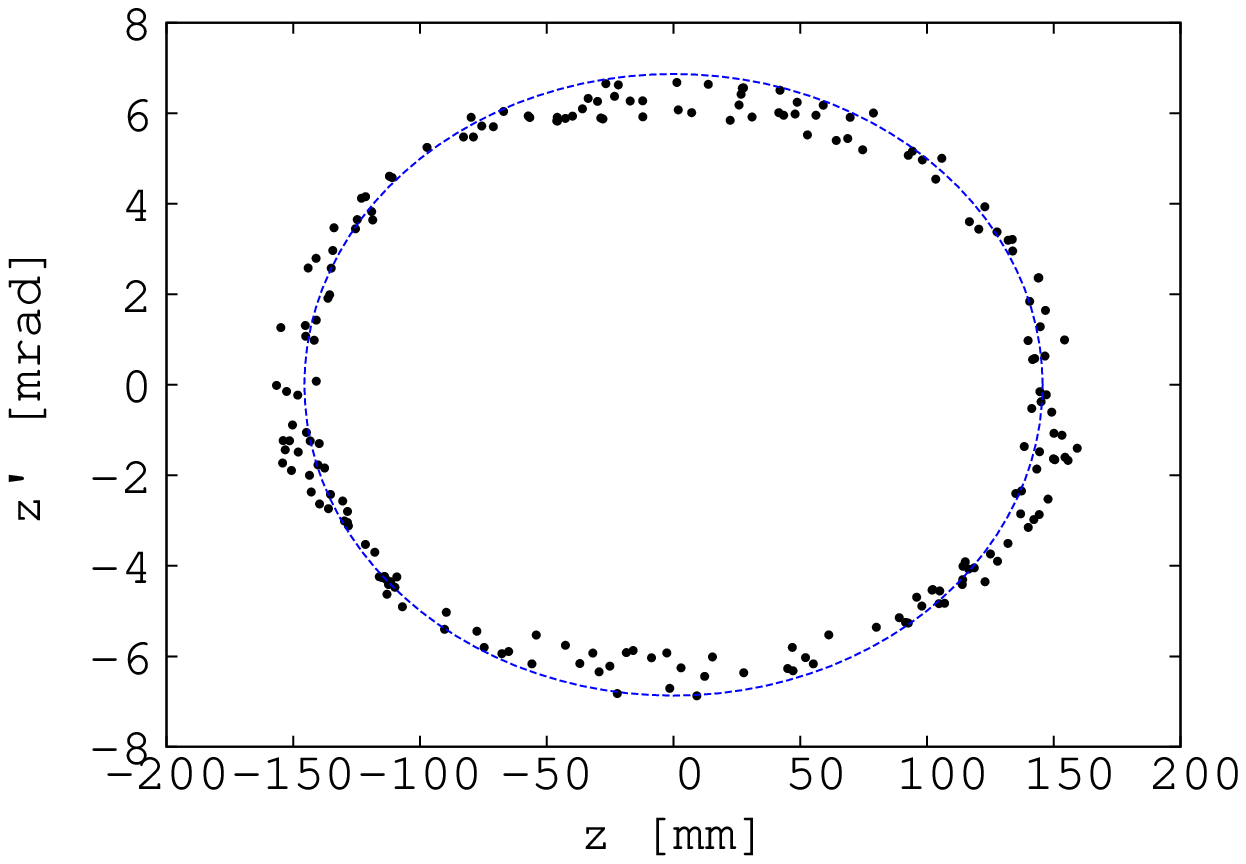}
	\caption{Stable motion in the vertical Poincare map for maximum initial amplitude over 100 turns for $p_0$, in PyZgoubi. The ellipse shows a 1~mm.rad unnormalized emittance.}
	\label{poincarrez-pyzgoubi}
   \end{minipage}
\end{figure}

To compare the models in the Runge-Kutta tracking in JBT and PyZgoubi, we show the vertical magnetic field for $p_0$ closed orbit in Fig.~\ref{bz} (top plots). The corresponding beta-functions at $p_{0}$ are shown in Fig.~\ref{beta}. Maximum amplitudes with a stable motion at $p_0$ over 100 turns are presented for horizontal and vertical planes in Fig.~\ref{poincarrex-pyzgoubi} and in Fig.~\ref{poincarrez-pyzgoubi}, respectively, all computed with PyZgoubi. 
There is a good agreement between the two codes, verifying the model and the results for optics and transverse acceptance. There is a small vertical beta-function mismatch observed between the straight part and the circular part in PyZgoubi, due to the difference in vertical phase advance between the JBT code and PyZgoubi in the straights. Even with this mismatch, a similar unnormalized transverse acceptance of about 1~mm.rad is found in both planes in Pyzgoubi, verifying this important long-term stability result.

As the pion distribution downstream of the target and the horn is uniform as a function of momentum~\cite{nustorm-horn-ao}, the beam intensity for parent pion and daughter muon beams in the nuSTORM ring is proportional to momentum acceptance. This means the FFAG lattice presented here offers a significant benefit in terms of flux. The FODO ring solution using an ideal lattice without imperfections showed 65\% beam transmission within $\approx\pm$8\% momentum acceptance~\cite{ao-thesis}, which is limited by the chromatic betatron mismatch between the arc and the straight section. Moreover once the realistic lattice is modelled, further reduction in FODO transmission is expected due to the effect of resonance crossing. The current FFAG solution double the momentum acceptance since it reaches $\approx\pm$16\% value with almost 100\% transmission. As an FFAG remains achromatic for all momenta the effect of chromatic mismatch and resonance crossing can be neglected. Although the FODO on-momentum transverse acceptance cannot be exceeded by the FFAG, it has a strong off-momentum dependence, while FFAG can preserve a large transverse acceptance for the entire momentum band. 

\begin{figure}[h!]
	\begin{center}
		\includegraphics[width=12cm]{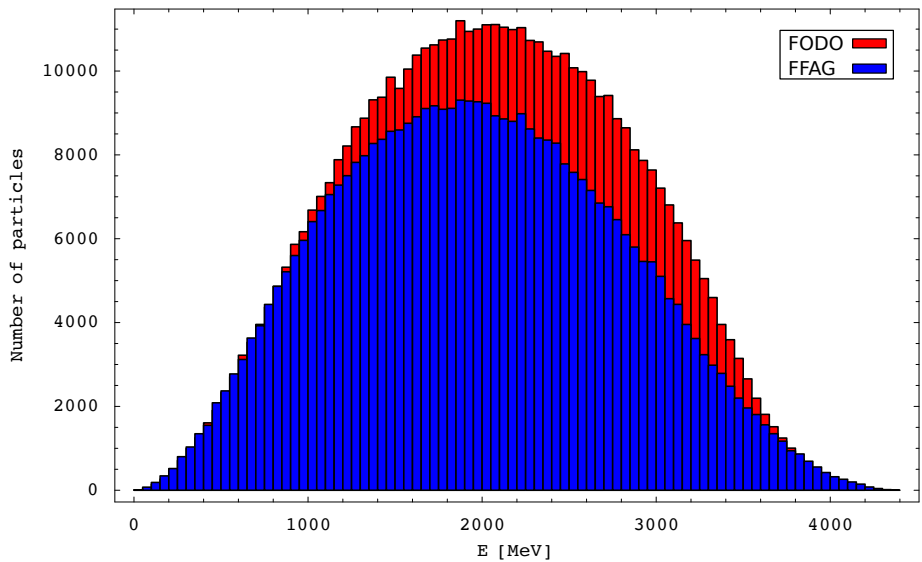}
		\caption{Relative flux of muon neutrinos as a function of neutrino energy from the triplet FFAG ring (blue histogram) relative to the one from the FODO ring (red histogram) assuming equal number of stored muons.}
		\label{flux}
	\end{center}
\end{figure}

However, there are also effects, which reduce the neutrino flux produced in an FFAG lattice, which are not present in the FODO lattice. Both are related to the unavoidable presence of bending magnets in the FFAG straight, which is necessary to preserve zero chromaticity. Firstly this introduces the scallop angle which periodically changes the direction of the beam in the straight section affecting not only the neutrino flux intensity at the detector, but also the shape of its spectrum. In order to minimise this effect the minimal scallop angle allowed by the optics was chosen and the triplet was selected as a focusing structure in the production straight allowing for long drifts pointing directly towards a neutrino detector. Fig.~\ref{flux} shows a relative flux as a function of neutrino momentum from the triplet FFAG ring relative to the one from the FODO ring assuming equal number of stored muons in both cases. The scallop effect both in the intensity and spectrum can be clearly seen. 
Secondly the presence of dispersion in the production/injection section reduces the efficiency of the muon storage. Both effects are under study and are presently estimated to reduce the effective flux in the near detector by $\approx$20\%. Coupled with a doubled momentum acceptance, these estimations promise the FFAG can give better performances than the FODO solution with respect to the neutrino flux production.

\section{Summary}
\label{sec:summ}

The nuSTORM project addresses essential questions in the neutrino physics, in particular by offering the best possible way to measure precisely the neutrino cross sections and by allowing the search for light sterile neutrinos. It would also serve as a proof of principle for the Neutrino Factory and can contribute to the R\&D for future muon accelerators.

We have shown here that a FFAG decay ring can meet the requirements of nuSTORM in terms of flux, so that the design can achieve a large momentum acceptance (3.8~GeV/c$\pm$16\%) and large transverse acceptance (1~mm.rad in both planes). This lattice is the most advanced design in the FFAG developments so far with 3 different types of cells matched and different field laws in the ring. The FFAG racetrack ring has some good features like increasing significantly the momentum acceptance simultaneously keeping large transverse acceptance, which should improve performance of nuSTORM facility.

The next step is a full comparison between FODO and FFAG solutions regarding neutrino flux performance, which is now made possible by this work. The final comparison of both type of lattices 
requires to perform 
the full study of muon capture efficiency followed by the multi-turn tracking in both rings including realistic lattices with imperfections and calculating the resulting neutrino flux. 

\section{Acknowledgements}
This work was supported by the US Department of Energy (DOE) and the UK Science and Technology Facilities Council (STFC) in the PASI (Proton Accelerators for Science and Innovation) framework, including the grants ST/G008248/1 and ST/K002503/1. Authors wish to acknowledge this support.

\end{document}